\documentstyle[12pt,aaspp4]{article}

\begin{document}

\title{Characterizing the QPO Behavior of the X-ray Nova
XTE~J1550--564} \author{Ronald A. Remillard\altaffilmark{1}, Gregory
J. Sobczak\altaffilmark{2}, Michael P. Muno\altaffilmark{1}, and
Jeffrey E. McClintock\altaffilmark{3}}

\altaffiltext{1}{Center for Space Research, MIT, Cambridge, MA 02139;
rr@space.mit.edu, muno@space.mit.edu}
\altaffiltext{2}{Harvard University, Astronomy Dept., 60 Garden
St. MS-10, Cambridge, MA 02138}
\altaffiltext{3}{Harvard-Smithsonian Center for Astrophysics, 60
Garden St. MS-3, Cambridge, MA 02138; jem@cfa.harvard.edu}

\begin{abstract} 

For all 209 RXTE observations of the X-ray nova XTE~J1550--564 during
its major outburst of 1998-1999, we have analyzed the X-ray power
spectra, phase lags, and coherence functions.  These observations
constitute one of the richest and most complete data sets obtained for
any black hole X-ray nova. The phase lags and coherence measures are
used to distinguish three types of low-frequency QPOs (one more than
those reported by Wijnands, Homan, \& van der Klis 1999). For the most
common type (``C''), the phase lag is correlated with both the QPO
frequency and the amplitude. The physical significance of the QPO
types is evident in the relationships between QPO properties and the
apparent temperature and flux from the accretion disk. There is also a
clear pattern in how the QPO types relate to the presence of
high-frequency QPOs.  In general, both the amplitude and the $Q$ value
($\nu / FWHM$) of low-frequency QPOs decrease as the high-frequency
oscillations increase in frequency (100 to 284 Hz) and in $Q$
value. We speculate that the antagonism between low-frequency and
high-frequency QPOs arises from competing structures in a perturbed
accretion disk. However, we find that the frequencies of slow ($< 20$
Hz) and fast ($ > 100$ Hz) QPOs are not correlated. In addition, we
encounter systematic problems in attempting to reliably compare the
QPO frequencies with broad features in the power continuum, since
there are a variable number of features or spectral breaks in the
power spectra.  These results cast some doubt on the reported global
relationship between QPOs from neutrons stars and those from black
hole systems.

\end{abstract}


\section{Introduction}

XTE~J1550--564 is an X-ray nova and black hole candidate discovered
with the All Sky Monitor (ASM; \cite{lev96}) on board the {\it Rossi
X-ray Timing Explorer} (RXTE) in 1998 September (\cite{smi98}).  A
bright and prolonged outburst from this source lasted $\sim 250$ days,
until 1999 May. During this time RXTE performed a series of pointed
observations on a nearly daily basis. Extensive spectral and timing
studies of this source have been performed using these data.  The
source has been observed in the ``very high'', ``intermediate'', and
``high/soft'' spectral states of black hole X-ray novae (\cite{sob99};
\cite{wij99a}; \cite{sob00b}; \cite{hom01}).  The X-ray spectrum
usually displays two components that can be well modeled as thermal
emission from an accretion disk (k$T \sim 1$ keV) and a power-law
component that extends to at least 150 keV (\cite{sob00b}).  A summary
of the observations, the values of the spectral parameters, and the
integrated measures of the disk flux and the power-law flux are
tabulated for all 209 observations during the 1998-1999 outburst in
Sobczak et al. (2000b)\nocite{sob00b}. Additional X-ray outbursts with
significantly weaker maxima and shorter duration were observed with
the ASM in 2000 April - June (\cite{smi00}) and again in 2001 February
- March (\cite{tom01}).

Optical observations of XTE~J1550--564 have revealed the binary period
of 1.54 days (\cite{jai01}). However, the optical mass function is not
yet measured, and the source remains a black hole candidate on the
basis of its X-ray timing and spectral characteristics.  Radio
observations have revealed evidence for a relativistic jet associated
with the large X-ray flare of 1998 September 19 (\cite{han01}). The
radio flux subsequently decayed away while the X-ray source remained
in the ``very high'' state.  Weaker radio flux with an inverted
spectrum was seen while the source was in the X-ray ``low-hard'' state
during the outburst of 2000 (\cite{cor01}), and this has been
interpreted to indicate the presence of a steady, compact jet at that
time.

In some observations during the 1998-1999 outburst, the X-ray power
spectra of XTE~J1550--564 exhibit quasi-periodic oscillations (QPOs) at
both high frequencies (HFQPOs; $\sim100$--285~Hz; \cite{rem99};
\cite{hom01}) and low frequencies (LFQPOs; 0.08--18~Hz; \cite{cui99a};
\cite{sob00a}).  The LFQPOs are sometimes particularly strong, with
peak to trough ratios as high as 1.5. (see Fig. 1 of Sobczak et
al. 2000b). The latter study shows that the frequency and amplitude of
QPOs below 20 Hz are correlated with both the power-law and the disk
components in the X-ray spectrum. These QPOs are observed only when
the power-law component contributes more than 20\% of the 2--20~keV
flux; at the same time, the QPO frequency is directly correlated with
the amount of disk flux (unabsorbed) seen in the 2--20 keV band
(\cite{sob00a}).

Recent efforts to understand LFQPOs have turned attention to the phase
lags associated with these oscillations and their harmonics. The
analysis technique uses Fourier cross spectra to measure both the
phase lags and the coherence parameter (versus frequency) between
selected ``soft'' and ``hard'' X-ray energy bands. The results
obtained for the first 13 RXTE observations of XTE~J1550--564 showed
unexpected phase lags in which the variations in hard X-ray preceded
those in soft X-rays (\cite{cui00}). Wijnands, Homan, \& van der Klis
(1999)\nocite{wij99a} used 14 observations of XTE~J1550--564 near the
end of the 1998-1999 outburst to begin organizing LFQPO properties in
terms of two different QPO types. Their first type (``A'') is broad
and exhibits a phase lag in soft X-rays; their second type (``B'') is
narrower and exhibits hard lags in the strongest feature but soft lags
in the harmonics.

In the present paper, we begin with an investigation of the phase lags
associated with LFQPOs for all 209 RXTE observations of XTE~J1550--564
that cover the 250~day outburst of the source during 1998-1999. We
find three QPO types, one in addition to those reported by Wijnands et
al.~(1999)\nocite{wij99a}. We then assess the correlations between the
phase lags, other properties of LFQPOs, the X-ray spectral parameters,
and also the properties of HFQPOs.

\section{Observations and Data Analysis}

Most of the PCA data were accumulated with $125\mu$s resolution in
``single bit'' mode, providing two energy channels corresponding
roughly to 2--6 and 6--13~keV, respectively (modes
SB\_125us\_0\_17\_1s \& SB\_125us\_18\_35\_1s).  Above 13~keV, ``event
mode'' was used with 16 energy channels and $16\mu$s time resolution
(mode E\_16us\_16B\_36\_1s).  In this investigation the single-bit
data and the event-mode data were each combined into a single channel,
resulting in two energy bands covering 2--13~keV and 13--30~keV,
respectively.  We chose 13-30 keV for the ``hard'' band in order to
isolate the power-law component and to select the channel (among the 3
choices) where the QPOs are often the strongest, in terms of percent
amplitude. Different data modes forced us to use alternative
channel boundaries for five of the observations, and the details are
given in the footnotes to Table~1.

The power density spectra (PDS) were computed for XTE~J1550--564 in
the sum energy band (i.e. 2--30 keV), as described in Sobczak et
al.~(2000a)\nocite{sob00a}.  The power spectra and cross spectra,
described below, were computed for every 256 s data segment.  Then,
for each observation we averaged together the results from the
individual segments. Finally, the PDS for each observation are
averaged in logarithmically increasing intervals of frequency ($\nu$)
prior to their display in units of log power density versus log $\nu$.

We compute the phase lag and coherence function between the 2--13~keV
and 13--30~keV energy bands as described in Bendat \& Piersol (1986)
\nocite{ben86} and Vaughan \& Nowak (1997)\nocite{vau97}.  The
coherence functions reported here include a correction for the effects
of Poisson noise.  When the coherence value is very low, then the
measured phase lag cannot be trusted (see \cite{now99}). To avoid
showing meaningless results in Figs. 1--3, we refrain from plotting
the phase lags and coherence at high frequency when the statistical
uncertainty in the coherence function is near unity.

We detected LFQPOs (i.e. below 20 Hz) in 72 of the 209 PCA observations
during 1998-1999; the results of our analysis of these QPOs are given
in Table 1. For a majority of LFQPOs, the profiles show three peaks
with frequencies in the ratio 1:2:4. In the final analysis we fit for
the amplitude and FWHM of each peak individually, using Lorentzian
functions. However, we used only one free parameter for the QPO
frequency, which forced the central frequencies of the three peaks to
be harmonically related, as indicated above.  There were only 4 cases
where the fit residuals indicated a slight yet significant departure
from this harmonic relation (Obs. \# 31, 43, 156, \& 162; see Table
1). These results confirm that the peaks are harmonically related, and
we suggest that the four exceptions are likely caused by the evolution
of the QPO waveform during the observation in question.

\subsection{The ``Fundamental'' QPO Feature}

The top panels of Figures~1--3 show some representative power density
spectra.  Here we focus on the triple-peaked QPO profiles, which occur
for most of the 72 cases reported in Table 1. The central peak with
frequency $\nu_{\rm o}$ strongly dominates over the satellite
features, which occur at 0.5$\nu_{\rm o}$ and 2$\nu_{\rm o}$.  If the
lowest frequency, 0.5$\nu_{\rm o}$, is the fundamental frequency of
the system, then the other two peaks correspond to the first and third
harmonics, while there is no apparent feature at the second harmonic
(1.5$\nu_{\rm o}$).  We derived upper limits for a peak at the second
harmonic for five power spectra that have pronounced triple-peaked
QPOs (Obs. \# 52, 53, 54, 159, \& 163; see Table 1) by adding another
Lorentzian peak at 1.5$\nu_{\rm o}$ to the fitting model. No
significant new detections were obtained.  The statistical
fluctuations produced candidate features with amplitudes
(i.e. integrated rms power normalized to the mean count rate) that
ranged from $<$ 0.1\% to $<$ 0.7\%. The average of the five
amplitudes, 0.4\%, is very small compared to the corresponding
amplitudes of the observed peaks at 0.5$\nu_{\rm o}$ and $\nu_{\rm
o}$, which are greater by a factor of 5 and 10, respectively. Our
failure to detect a second harmonic component suggests that
0.5$\nu_{\rm o}$ may not be the fundamental frequency of the system.

A second and perhaps more compelling reason to focus attention on the
strongest QPO feature is that there are time intervals in which the
feature at 0.5 $\nu_{\rm o}$ gradually fades below the detection
threshold, while the relative strength of the other peaks appears to
remain constant. For example, consider observations \# 3--14 in Table
1.  The feature at 0.5 $\nu_{\rm o}$ falls in and out of detectability
while the features at $\nu_{\rm o}$ and $2\nu_{\rm o}$ are
continuously detected.  This provides motivation to focus attention on
the strongest QPO feature.  On the other hand, if we were to
redefine the fundamental QPO as the one detected with the lowest
frequency, then we would find large variations in the relative
amplitudes of the harmonics, and frequency discontinuities would
appear in the correlation plots given below.

We therefore adopt an observational strategy and {\it presume} that
the strongest QPO feature in XTE J1550--564 is the ``fundamental''.
From that perspective, the first harmonic is almost always present
with a strength (relative to the fundamental) of roughly 0.1 to 0.5,
similar to that of GRS1915+105 (see \cite{mun99}).  In 17\% of the
observations we fail to detect a peak at 0.5$\nu_{\rm o}$; however,
its amplitude rises to 60\% relative to the QPO at $\nu_{\rm o}$ on
some occasions (e.g. 1998 October 15).  When we do detect the
subharmonic, its $Q$ value ($Q=\nu/ FWHM$) is almost always less than
the $Q$ of the fundamental. Our identification of the dominant QPO
peak with the fundamental frequency imposes a serious challenge to
identify a physical mechanism that can generate X-ray power at
0.5$\nu_{\rm o}$; however, we regard this consequence as less of a
problem than the alternative course. In the format for Table 1, we
first list the LFQPO fundamental, and the frequency of any HFQPO is
given in the last column on that line. The subsequent two lines report
the properties of the first harmonic and then the subharmonic, if the
detections are above 3 $\sigma$.

\subsection{Broadband Features in the Power Continuum}

The relationship between QPO frequencies and the frequencies of other
broad features or spectral breaks in the power continuum have been
investigated by Wijnands \& van der Klis (1999)\nocite{wij99b} and
Psaltis, Belloni \& van der Klis (1999)\nocite{psa99}. They report
frequency correlations and argue that the accreting black holes and
neutron stars exhibit essentially similar patterns of PDS
variability. The results were interpreted in terms of a relativistic
precession model that applies to compact objects of both types
(\cite{ste99}).

We have investigated this topic for the case of XTE J1550--564, using
all of the observations during the 1998-1999 outburst. There are broad
features in the power continuum, with $Q < 1$, and we have attempted to
measure these by identifying either the breaks in the PDS (assuming
power law functions for the frequency intervals between the breaks), or
by fitting for peaks in the ($\nu \times PDS_{\nu}$) vs. $\nu$ plane. In
both cases, we find systematic problems that prevent us from drawing
any firm conclusions. We have also investigated the relationship
between LFQPOs and HFQPOs, both of which generally pertains to PDS
features with $Q > 2$, and we report those results in Section 3.2
below.

In our attempts to define breaks in the power continuum, we find that we
cannot obtain a satisfactory fit for all of the power spectra using
one mathematical model.  Many of the power spectra during the early
phase of the outburst (see Fig. 2 in \cite{cui99a}) are well fit above
0.1 Hz by a model consisting of two continuum breaks and a QPO with
one harmonic. However, in other observations, e.g during 1998
September 15-22 (see Fig. 1 in \cite{rem99}), three to five continuum
breaks are required.

The broad features in the power continuum of XTE~J1550--564 are most
apparent as peaks in the ($\nu \times PDS_{\nu}$) vs. $\nu$ plane
(\cite{psa99}).  We find that there are generally one to three broad
peaks per observation, and the profiles of these peaks often deviate
from simple mathematical models. The LFQPOs may or may not lie near the
center of one peak. Further progress on this analysis topic requires
additional efforts to solve the modeling problems and to devise
selection criteria to guide the comparison of broad features and QPO
detections.  Such tasks are beyond the scope of this paper, and we are
presently unable to derive useful conclusions regarding the
relationship between the QPO frequencies and the broadband peaks and
breaks in the power continuum for XTE~J1550--564.

\section{Results}
\subsection{QPO Types}

We find three fundamental types of phase lag behavior for
XTE~J1550--564. Figures 1 and 2 show representative power density
spectra, phase lags, and coherence functions for the three types of
QPOs.  Types A and B correspond to those identified previously by
Wijnands et al. (1999)\nocite{wij99a}, while type C QPOs (narrow, 
with soft phase lags) appear during the first half of the outburst in
observations that they did not analyze. In addition to types A, B and
C, anomalous QPOs were observed on two occasions, and their power
spectra are shown in Fig. 3.

Type A timing behavior is characterized by a broad QPO ($Q \sim 2$--3)
near 6~Hz that appears to be a superposition of the fundamental and
the first harmonic (Fig.~1).  The integrated rms amplitude of the QPO
features is a few percent.  The phase lags for type A behavior are
generally featureless except for a broad soft lag centered near the
QPO feature with $\vert \Delta \phi \vert \sim 1$~rad and poor
coherence ($< 50$\%). We identify four observations with type A QPO
behavior, all of which display HFQPOs in the range 270-284 Hz (see
Table 1).  There are six additional observations in which the QPO is
broad but the statistics are inadequate to establish a soft phase
lag. These are labeled ``A?'', and they include the last three QPO
detections (Table 1) in which the QPO is only evident above 6 keV.  Of
the six ``A?''  cases, an additional four have HFQPOs in the range
214-284 Hz. Typically, the HFQPOs have $Q \sim 10$, and they are maximally
detected by ignoring the PCA counts below 6 keV.

Type B QPO behavior is characterized in the 1998 October 20
observation (Fig.~1) by a narrow fundamental QPO feature ($Q \sim 10$)
at 5.5 Hz (see Fig.~1) with rms amplitude of 3.5\%.  Type B
observations display a hard lag ($\Delta \phi \sim 0.0$--0.4~rad) that
is slightly shifted from the fundamental feature, while there are soft
lags associated with the first and subharmonics.  The coherence of
the type B QPO is typically within 10\% of unity. Type B QPOs also
have strong subharmonics that are typically 30-60\% of the amplitude
of the fundamental. We identify 9~observations with type B QPO
behavior. Six of these additionally display HFQPOs in the
range 182-209 Hz. These HFQPOs are systematically lower
in frequency compared to the type A group, they are broader in width
($Q \sim 5$), and their amplitude is more broadly distributed with
X-ray photon energy.

We introduce type C timing behavior in Fig.~2. The 1998 October 10
observation (Obs. \#43) also happens to be one of the four previously
noted cases in which the harmonics are slightly shifted from the
expected positions, shown with dashed lines in the left panel of
Fig. 2.  The slight frequency offsets are barely discernible in
logarithmic units.  Type C QPOs occur primarily during the first half
of the outburst, when the source is in the very high state and the
power-law component contributes $\gtrsim 75$\% of the 2--20~keV flux
(\cite{sob00b}).  Type C timing behavior is characterized by sharp ($Q
\gtrsim 10$) fundamental QPO features with a range of rms amplitudes
from 3--16\%.  The first harmonic is present, and the first
subharmonic is usually observed as well.  The phase lags for type C
behavior are typically modest, with the fundamental exhibiting soft
lags $\vert \Delta \phi \vert \lesssim 0.4$~rad.  The subharmonic
usually has a small soft phase lag of the same magnitude as the
fundamental, whereas, the first harmonic has a hard lag, which can be
several times larger than the soft lag displayed by the fundamental.
The coherence of the fundamental is high $\sim$85--95\%, while the
coherence of the first and subharmonic is typically closer to
75--80\%.  We identify 51~observations with type C QPO behavior. Only
five of these display HFQPOs, and their frequencies are all below 170 Hz.

Among the 51 observations of type C QPOs, there are 5 cases that are
distinguished by weak harmonics and a double-peak structure within the
profile of the fundamental. An example is shown in Fig. 2. These five
cases are labeled ``C\'{ }'' in Table 1; they all occur during the two
days (1998 September 20-21) immediately following the 6.8~Crab flare
in the XTE~J1550--564 light curve. If we divide any of these
observations into smaller time intervals, we find a single narrower QPO,
suggesting that the QPO shifts in frequency during the
observation. The harmonics are difficult to discern for type C\'{ }
QPOs; nevertheless, there is significant structure in the phase lag
near the harmonic frequencies (see the dotted vertical lines in
Fig. 2).  Apart from their PDS profiles, the C\'{ } QPOs exhibit
properties that overlap with the other type C cases: frequencies
from 6--10~Hz, rms amplitudes of 3--7\%, and phase lags $\sim
-0.2$~rad. Therefore, in our judgment, the C\'{ } QPOs are best
described as type C QPOs with frequency shifts that are probably
related to the system's recovery from the huge X-ray flare that
preceded those observations.

The properties of the three fundamental types of LFQPOs in
XTE~J1550--564 are summarized in Table~2. There are two observations
that do not conform to these QPO categories, and they are shown in
Fig.~3.  The first anomalous QPO occurred during the 6.8~Crab flare on
1998 September 19 (\cite{sob99}). This observation produced a 13~Hz
QPO with an integrated rms amplitude of 1\%, a large soft lag of
$-0.9\pm0.2$~rad, and a coherence of $80\pm20$\%.  The power-law
dominates the spectrum during this flare, with the disk contributing
only $\sim3$\% of the 2--20~keV flux. The magnitude of the soft phase
lag is consistent with type A behavior; however, the presence of a QPO at 183
Hz resembles type B cases, but the LFQPO frequency and the low
fractional contribution of the disk flux is similar to type C
behavior. The second anomalous QPO occurred on 1999 March 2.  This LFQPO
(18~Hz) is observed at a much higher frequency than the others, while
the coherence at all frequencies is very low.  It is significantly
sharper than the other LFQPOs ($Q \sim 19$) and has a very small rms
amplitude (0.5\%) with no harmonics.

\subsection{Correlations with the Accretion Disk and High-Frequency QPOs}

We use the measured QPO properties (Table~1) to investigate
the relationship between the phase lag and the frequency and amplitude
of the fundamental feature.  The results are shown in Fig.~4, where we
use different plotting symbols to identify the QPO type.  The
relationships clearly appear more organized when the QPO type is
distinguished. For type C QPOs, as the phase lag decreases below zero
(i.e. increasing soft lag), the QPO frequency increases and the
amplitude decreases.

In Fig.~5, we examine how the QPO parameters vary in response to the
observed spectral changes in the X-ray component attributed to the
accretion disk. In the left panels of Fig. 5, we plot the LFQPO
parameters versus the disk color temperature, using the spectral
parameters given in Table 2 of Sobczak et al. (2000b)
\nocite{sob00b}. Again, the results appear more organized when the QPO
type is distinguished. In the central panels in Fig. 5 we plot the QPO
parameters versus the unabsorbed disk flux, here restricting the
integration over photon energy to the range 2-20 keV, as given in
Table 3 of Sobczak et al. (2000b) \nocite{sob00b}. There is a very
striking linear correlation between the QPO frequency and the amount
of disk flux that falls within the bandwidth of the PCA instrument. In
the right panels, the QPO parameters are plotted versus the bolometric
disk flux: $f_{dbb} = 2.16 \times 10^{-11} N T^4$ erg cm$^{-2}$
s$^{-1}$, where $N$ and $T$ are the normalization and color
temperature, respectively, for the accretion disk, as derived from the
X-ray spectral analyses. Compared to the central panels in Fig. 5, the
increased scatter in the right panels is probably a consequence of
the fact that most of the disk flux is below 2 keV. Thus the
calculation of the total disk flux requires extrapolation of the
spectrum well below the threshold of the PCA instrument, and even
small statistical or systematic errors in the temperature measurement
may cause significant errors in the calculation of the bolometric disk
flux. Alternatively, we cannot exclude the possibility that the
thermal X-ray component actually diverges from the simple disk
blackbody model at low photon energies.

Overall, the frequency of type C QPOs is most correlated with the disk
flux, while the QPO amplitude and phase lags show slightly greater
correlation with the disk temperature (see Fig. 5).  We further note
that while the sequence of QPO types C $\rightarrow$ B $\rightarrow$ A
is associated with increased total disk luminosity, the disk's
contribution to the total flux decreases along this sequence, and the
total luminosity is greatest when we observe the huge flare of 1998
September 19 (with an anomalous QPO) and the subsequent decay with
type C\'{ } QPOs.  We have plotted the QPO parameters versus the
spectral properties of the power-law component (not shown), and the
results produce only rough correlations, with increased scatter
compared to the left and right panels of Fig. 5.

In Fig. 6, we examine the phase lag of LFQPOs versus the fraction of
the unabsorbed flux contributed by the accretion disk.  The disk
fraction is calculated in two ways, analogous to the central and right
panels of Fig. 5. In the left panel of Fig. 6, the flux for both
spectral components is integrated over the range 2--20 keV.  In the
right panel we consider the entire disk-blackbody spectrum, and we
integrate the power-law component over the range 1--30 keV. The intent
of the latter method is to track the bolometric flux from each
spectral component, but the calculations are subject to increased
systematic uncertainty, as noted with respect to Fig. 5.  The results
show that each types of QPO occupies a distinct region in Fig.~6. 
The situation is clearest in the left panel where the
three types of QPOs segregate into distinct regions:
the C type QPOs occur when the fraction
$\lesssim 0.26$, type A QPOs occur when the disk fraction
$\gtrsim 0.37$, and type B QPOs occur when the disk
fraction lies between these values. We further note that QPOs are not
detected for the majority of observations in which the disk fraction
exceeds 0.4 (or 0.5 for bolometric fraction), and there is only a one
(peculiar) QPO detection when the disk fraction exceeds 0.8.

Finally, for the 20 observations with HFQPOs listed in Table 1, we plot
the frequencies of HFQPOs versus the LFQPO fundamental in the top panel
of Fig.~7; the frequencies appear to be uncorrelated.  A weak
correlation would be produced if we plot the LFQPO frequencies at $0.5
\nu_{\rm o}$ for types B and C only, but our reasons for not doing so
are given in Section 2.1. However, in the same plot we do see a clear
relation between LFQPOs and HFQPOs in that the symbol type changes
progressively with the frequency of the HFQPO. All of the high-Q
detections near 283 Hz are associated with broad, type A QPOs at 5-10
Hz, while the cluster of broader QPOs near 183 Hz mostly occur with
more narrow, type B QPOs at 5-7 Hz.  The trend toward an
anti-correlation in the $Q$ vales ($\nu / FWHM$) of HFQPOs and LFQPOs is
shown in the bottom plot of Fig.~7. We further note that the amplitude
of the fundamental feature decreases with types A and B (see Fig. 5),
when HFQPOs are frequently detected. Using the results given in Table
1, it can be seen that the LFQPO amplitudes are generally (19 of 20
cases) in the range 1--8\% when HFQPOs are detected, but when HFQPOs are
not detected the LFQPO amplitude is generally (43 of 52 cases) in the
range of 8--16\%. We conclude that the amplitudes of LFQPOs and HFQPOs
also appear to be anti-correlated in XTE~J1550--564.

\section{Discussion}

We have shown that the effort to distinguish QPO types by their
measured phase lag and coherence is useful in sorting out the
properties of LFQPOs; furthermore, it provides insight into the
relationship between LFQPOs and HFQPOs in XTE~J1550-564.  The QPO
types also help to establish certain strong correlations between the
properties of the type C oscillations and the spectral properties of
the accretion disk. These correlations, particularly the one in the
top-central panel of Fig. 5, also provide assurance that the spectral
deconvolution of the two spectral components (i.e. thermal and power
law) using the RXTE PCA instrument is physically meaningful, and this
should encourage more sophisticated disk models that may allow us to
interpret the physical conditions in the inner disk without the
uncertainties inherent in the simple multi-temperature disk model.

Coherence functions and phase lags are expected to provide powerful
constraints on the X-ray power-law component of accreting black holes
when interpreted with a physical model for Comptonization.  In the
extended corona model (e.g. \cite{miy88}; \cite{hua96}; \cite{kaz97};
\cite{now99}), the hard photons undergo more scatterings in order to
reach higher energies and therefore lag behind the soft photons in the
response to flares from the inner accretion disk. The hard lag is
directly related to the photon diffusion time scale through the
corona, which scales logarithmically with photon energy
(\cite{hua96}), in agreement with some observations (\cite{cui97};
\cite{now99}). This scenario has been invoked to constrain the
structure of an extended but inhomogeneous Compton corona
(\cite{kaz99}).  However, problems accommodating large observed phase
lags have prompted alternative models in which the lags signify the
spectral and temporal evolution of the parent mechanism of
Comptonization. This may occur, e.g., with magnetic loop structures in
the disk (\cite{pou99}), where the hot plasma associated with magnetic
instabilities may produce secondary effects such as disk heating and
Comptonization.

The topic of QPO phase lags and coherence measures is especially
difficult because one must deal with an unknown oscillation mechanism.
When LFQPOs are detected there is always a substantial X-ray power
law, but the relationship between the QPOs and the X-ray spectral
components is very complicated, as shown herein and elsewhere
(e.g. \cite{mar99}; \cite{mun99}; \cite{sob00a}).  The model for
Comptonization via bulk motion flow does predict a correlation between
disk flux and QPO frequency (\cite{tit98}); however, the applicability
of this model to LFQPO below 20 Hz is questionable, and the measured
soft lags appear to be problematic. A proposal to explain QPO phase
lags in GRS~1915$+$105 has been offered by Nobili et
al. (2000)\nocite{nob00}, who invoke radial oscillations in a
transition boundary between the disk and an inner Comptonizing
region. However, their model does not accommodate the wealth of
observational evidence presented herein and in similar RXTE studies.

It is clear, e.g. in Figs. 1-3, that the plots of phase lags and
coherence functions show features that coincide with the profiles of
the QPOs and harmonics.  The contrast between the QPOs and the
broadband power continuum, along with the sign reversals in the QPO
phase lags, might signify less about Comptonization and more about the
details of the QPO waveform. Similar sign reversals in phase lags were
also reported for LFQPOs in the microquasar and black hole candidate
GRS~1915$+$105 (\cite{cui99b}; \cite{rei00}; \cite{mun01}).  Further
insights regarding the QPO waveforms and their physical origin might
come from "QPO-folding" analyses such as those undertaken by Morgan,
Remillard, \& Greiner (1997)\nocite{mor97}.

HFQPOs are coincident with almost all type A and B LFQPOs and rarely
with QPOs of type C. The frequencies of the fast and slow QPOs in
XTE~J1550-564 do not appear to be correlated, casting some doubt on
the relationship proposed to unite the QPOs in both neutron-star and
black-hole systems (\cite{psa99}).  On the other hand there does
appear to be evidence of a connection between LFQPOs and HFQPOs in the
sense of an anti-correlation between their amplitudes and $Q$ values.
These comparisons are bounded by the disappearance of all QPOs when
the disk contributes more than $\sim 80$\% of the flux. On the other
hand there is a low probability of detecting HFQPOs when the disk
contributes less than $\sim 30$\% of the total flux {\it and} the
LFQPO amplitude is above 8\%. Taken as a whole, our results suggest
that both LFQPOs and HFQPOs arise in the inner accretion disk, that
the power-law component is associated with the excitation of these
oscillations, and that there is some mutual antagonism in the
structures that are responsible for HFQPOs and LFQPOs,
respectively. The accretion-ejection instability in magnetized disks
(\cite{tag99}) is one example of a mechanism that could cause such an
effect.  It is proposed that LFQPOs occur when the instability forms a
standing spiral wave that transfers energy from the inner disk to
larger radii. As the amplitude of the wave increases, it is
conceivable that the LFQPO amplitude would increase while energy
depletion in the inner disk might weaken the HFQPO.

\acknowledgements We thank E. Morgan, J.~Homan, and R.~Wijnands for
helpful discussions.  Partial support for R.R. and M.M.  was provided
by the NASA contract to M.I.T. for RXTE instruments, and for J.M. and
G.S. by NASA Grant NAG5-10813.

\begin{deluxetable}{lccccccccc}
\scriptsize 
\tablewidth{0pt}
\tablenum{1}
\tablecaption{Low-Frequency QPO Parameters for XTE~J1550--564:  
Fundamental and Harmonic Frequencies}
\tablehead{
 \colhead{Obs} & \colhead{Date} & \colhead{MJD\tablenotemark{a}} &
 \colhead{Type} & \colhead{QPO $\nu$\tablenotemark{b}} & 
 \colhead{Amplitude\tablenotemark{c}} & \colhead{Q\tablenotemark{d}} &
 \colhead{Phase Lag\tablenotemark{e}} & \colhead{Coherence\tablenotemark
 {f}} & \colhead{HFQPO\tablenotemark{g}} \cr
 \colhead{\#} & \colhead{(yymmdd)} & \colhead{} &
 \colhead{} & \colhead{(Hz)} & \colhead{(\% rms)} & \colhead{} &
 \colhead{(radians)} & \colhead{} &
 \colhead{(Hz)}
}
\startdata
1&  980907&  51063.72&  C &    0.081&  $13.1\pm1.4$&  16.2 &  
$0.10\pm0.22$&  $0.95\pm0.04$&  \nodata\nl
\nodata&  \nodata&  \nodata&  \nodata&    0.16 &  $ 11.5\pm0.9$&   7.9 &  
$0.11\pm0.20$&  $0.99\pm0.03$&  \nodata\nl
2&  980908&  51064.01&  C &    0.122 &  $12.3\pm0.7$ &   8.7 &  
$-0.03\pm0.01$&  $0.997\pm0.002$&  \nodata\nl
\nodata&  \nodata&  \nodata&  \nodata&    0.24 &  $ 8.0\pm0.6$&   5.2 &  
$0.08\pm0.02$&  $0.995\pm0.003$&  \nodata\nl
3&  980909&  51065.07&  C &    0.288 &  $14.6\pm0.9$ &   16.9 &  
$-0.01\pm0.01$&  $0.994\pm0.003$&  \nodata\nl
\nodata&  \nodata&  \nodata&  \nodata&    0.58 &  $ 5.1\pm0.4$&  12.5 &  
$0.10\pm0.03$&  $0.965\pm0.008$&  \nodata\nl
\nodata&  \nodata&  \nodata&  \nodata&    0.14 &  $ 3.8\pm0.5$&  6.8 &
$-0.04\pm0.03$&  $0.965\pm0.010$&  \nodata\nl
4&  980909&  51065.34&  C &    0.395 &  $13.5\pm0.8$& 13.6 &  
$-0.02\pm0.02$&  $0.985\pm0.004$&  \nodata\nl
\nodata&  \nodata&  \nodata&  \nodata&    0.79 & $ 6.1\pm0.6$ & 9.1 &
$0.09\pm0.05$&  $0.948\pm0.015$&  \nodata\nl
5&  980910&  51066.07&  C &    0.81 &  $14.3\pm0.5$&  13.1 &  
$0.05\pm0.02$&  $0.971\pm0.005$&  \nodata\nl
\nodata&  \nodata&  \nodata&  \nodata&    1.6&  $ 5.9\pm0.4$&  8.4 &  
$0.16\pm0.04$&  $0.888\pm0.017$&  \nodata\nl
6&  980910&  51066.34&  C &    1.03 &  $15.0\pm0.4$&  12.5 &  
$0.02\pm0.02$&  $0.962\pm0.007$&  \nodata\nl
\nodata&  \nodata&  \nodata&  \nodata&    2.1 &  $ 6.5\pm0.3$ & 7.5 &  
$0.11\pm0.04$&  $0.884\pm0.018$&  \nodata\nl
7&  980911&  51067.27&  C &    1.54 &  $15.1\pm0.4$ &  11.2 &  
$0.04\pm0.01$&  $0.971\pm0.004$&  \nodata\nl
\nodata&  \nodata&  \nodata&  \nodata&    3.1 &  $ 5.0\pm0.2$&  10.5 &  
$0.23\pm0.04$&  $0.801\pm0.021$&  \nodata\nl
8&  980912&  51068.35&  C &    2.38 &  $13.7\pm0.4$&  12.6 &  
$0.01\pm0.01$&  $0.963\pm0.004$&  \nodata\nl
\nodata&  \nodata&  \nodata&  \nodata&    4.8 &  $ 5.3\pm0.2$&   9.6 &  
$0.24\pm0.05$&  $0.765\pm0.026$&  \nodata\nl
\nodata&  \nodata&  \nodata&  \nodata&    1.2 &  $ 2.7\pm0.3$ &  4.8 &  
$0.02\pm0.06$&  $0.777\pm0.033$&  \nodata\nl
9&  980913&  51069.27&  C &    3.33 &  $12.8\pm0.3$&  15.1 &  
$-0.04\pm0.01$&  $0.974\pm0.003$&  \nodata\nl
\nodata&  \nodata&  \nodata&  \nodata&    6.7 &  $ 4.0\pm0.2$&  14.5 &  
$0.31\pm0.05$&  $0.759\pm0.030$&  \nodata\nl
\nodata&  \nodata&  \nodata&  \nodata&    1.7 &  $ 2.2\pm0.3$&  7.2 &  
$-0.06\pm0.05$&  $0.770\pm0.028$&  \nodata\nl
10&  980914&  51070.13&  C &    3.20 &  $13.1\pm0.3$&  16.0&  
$-0.03\pm0.01$&  $0.969\pm0.003$&  \nodata\nl
\nodata&  \nodata&  \nodata&  \nodata&    6.4 &  $ 4.2\pm0.2$ & 13.9&  
$0.30\pm0.04$&  $0.655\pm0.026$&  \nodata\nl
11&  980914&  51070.27&  C &    3.17 &  $13.0\pm0.3$\tablenotemark{j} &  21.1 &  
$-0.03\pm0.02$&  $0.959\pm0.004$&  \nodata\nl
\nodata&  \nodata&  \nodata&  \nodata&    6.3 &  $ 4.0\pm0.2$&  11.5&  
$0.38\pm0.05$&  $0.652\pm0.027$&  \nodata\nl
12&  980915&  51071.20&  C &    3.69 &  $12.2\pm0.2$\tablenotemark{j} &  14.8 &  
$0.00\pm0.01$&  $0.971\pm0.003$&  \nodata\nl
\nodata&  \nodata&  \nodata&  \nodata&    7.4 &  $ 3.8\pm0.2$&  12.9 &  
$0.34\pm0.05$&  $0.710\pm0.035$&  \nodata\nl
13&  980915&  51072.00&  C &    2.58 &  $14.7\pm0.7$&  23.5 &  
$0.08\pm0.04$&  $0.976\pm0.040$&  \nodata\nl
\nodata&  \nodata&  \nodata&  \nodata&    5.1 &  $ 5.2\pm0.3$&  17.2 &  
$0.35\pm0.06$&  $0.794\pm0.033$&  \nodata\nl
\nodata&  \nodata&  \nodata&  \nodata&    1.3 &  $ 2.8\pm0.3$&   7.9 &  
$0.22\pm0.09$&  $0.682\pm0.058$&  \nodata\nl
\tablebreak
14&  980916&  51072.34&  C &    4.02 &  $11.9\pm0.2$&  15.5 &  
$-0.07\pm0.02$&  $0.965\pm0.004$&  \nodata\nl
\nodata&  \nodata&  \nodata&  \nodata&    8.0 &  $ 3.3\pm0.2$&  14.6 &  
$0.29\pm0.05$&  $0.687\pm0.03$&  \nodata\nl
\nodata&  \nodata&  \nodata&  \nodata&    2.0 &  $ 2.6\pm0.3$&  11.1 &  
$-0.08\pm0.05$&  $0.763\pm0.031$&  \nodata\nl
15&  980918&  51074.14&  C &    5.75 &  $ 8.8\pm0.2$&   9.6&  
$-0.163\pm0.02$&  $0.927\pm0.007$&  \nodata\nl
\nodata&  \nodata&  \nodata&  \nodata&   11.5 &  $ 1.3\pm0.1$&  12.4 &  
$0.02\pm0.07$&  $0.638\pm0.055$&  \nodata\nl
\nodata&  \nodata&  \nodata&  \nodata&    2.9 &  $ 1.8\pm0.2$&  7.4 &  
$-0.14\pm0.05$&  $0.860\pm0.024$&  \nodata\nl
16\tablenotemark{*}&  980919&  51075.99&  ?&   13.2 &  $ 0.8\pm0.1$&   
5.0 &  $-1.02\pm0.12$&  $0.82\pm0.15$&  $183\pm4$\nl
\nodata&  \nodata&  \nodata&  \nodata&    4.9 &  $ 0.26\pm0.03$&  14.6&  
$-0.40\pm0.13$&  $0.97\pm0.21$&  \nodata\nl
17&  980920&  51076.80&  C\'{ } &    7.2 &  $ 6.6\pm0.1$\tablenotemark{j} &   5.0&  
$-0.22\pm0.02$&  $0.859\pm0.009$& \nodata\nl
18\tablenotemark{h}&  980920&  51076.95&  C\'{ } &    8.5 &  $ 4.8\pm0.1$\tablenotemark{j} &   
4.3 &  $-0.23\pm0.02$&  $0.883\pm0.009$&  \nodata\nl
19\tablenotemark{h}&  980921&  51077.14&  C\'{ } &    9.8 &  $ 3.2\pm0.1$&   
9.2&  $-0.17\pm0.02$&  $0.883\pm0.014$&  $169\pm5$\nl
\nodata&  \nodata&  \nodata&  \nodata&   19.6 &  $ 1.2\pm0.1$&   5.0&  
$0.07\pm0.16$&  $0.71\pm0.18$&  \nodata\nl
20\tablenotemark{h}&  980921&  51077.21&  C\'{ } &    7.1 &  $ 4.0\pm0.1$\tablenotemark{k} &   
5.4&  $-0.19\pm0.02$&  $0.875\pm0.009$&  $159\pm4$\nl
21&  980921&  51077.87&  C\'{ } &    6.3 &  $ 10.2\pm0.4$\tablenotemark{k} &  5.8 &  
$-0.15\pm0.02$&  $0.901\pm0.005$&  \nodata\nl
22&  980922&  51078.13&  C &    5.41 &  $9.7\pm0.1$\tablenotemark{j} &  11.3 &  
$-0.15\pm0.02$&  $0.927\pm0.005$&  \nodata\nl
\nodata&  \nodata&  \nodata&  \nodata&   10.8 &  $ 2.1\pm0.2$&  13.0 &  
$0.25\pm0.04$&  $0.692\pm0.037$&  \nodata\nl
\nodata&  \nodata&  \nodata&  \nodata&    2.7 &  $ 1.6\pm0.2$&   5.0 &  
$-0.13\pm0.03$&  $0.870\pm0.015$&  \nodata\nl
23\tablenotemark{i}&  980923&  51079.79&  C &  3.90 & $12.1\pm0.3$ &   
13.9 &  $-0.04\pm0.03$&  $0.898\pm0.012$&  \nodata\nl
\nodata&  \nodata&  \nodata&  \nodata&    7.8 &  $ 3.3\pm0.2$&  12.6 &  
$0.25\pm0.10$&  $0.521\pm0.079$&  \nodata\nl
24\tablenotemark{i}&  980924&  51080.08&  C &    3.87 &  $11.9\pm0.2$&  
14.3 &  $-0.10\pm0.02$&  $0.931\pm0.008$&  \nodata\nl
\nodata&  \nodata&  \nodata&  \nodata&    7.7 &  $ 3.1\pm0.2$& 13.3 &  
$0.42\pm0.10$&  $0.450\pm0.070$&  \nodata\nl
\nodata&  \nodata&  \nodata&  \nodata&    1.9 &  $ 2.1\pm0.2$&  6.9 &  
$0.07\pm0.06$&  $0.740\pm0.041$&  \nodata\nl
25&  980925&  51081.06&  C &    2.88 &  $12.7\pm0.2$&  10.7 &  
$-0.01\pm0.01$&  $0.978\pm0.003$&  \nodata\nl
\nodata&  \nodata&  \nodata&  \nodata&    5.8 &  $ 4.0\pm0.2$& 10.3 &  
$0.25\pm0.04$&  $0.724\pm0.020$&  \nodata\nl
\nodata&  \nodata&  \nodata&  \nodata&    1.4 &  $ 2.3\pm0.2$&  8.4 &  
$0.02\pm0.04$&  $0.784\pm0.23$&  \nodata\nl
26&  980926&  51082.00&  C &    2.72 &  $13.6\pm0.2$&  10.9 &  
$-0.01\pm0.01$&  $0.973\pm0.020$&  \nodata\nl
\nodata&  \nodata&  \nodata&  \nodata&    5.4&  $ 4.4\pm0.2$&  12.1 &  
$0.23\pm0.02$&  $0.810\pm0.015$&  \nodata\nl
\nodata&  \nodata&  \nodata&  \nodata&    1.4 &  $ 2.0\pm0.2$& 10.4 &  
$0.06\pm0.03$&  $0.769\pm0.019$&  \nodata\nl
27&  980927&  51083.00&  C &    2.66 &  $13.7\pm0.2$&  9.8 &  
$-0.01\pm0.01$&  $0.966\pm0.003$&  \nodata\nl
\nodata&  \nodata&  \nodata&  \nodata&    5.3 &  $ 4.5\pm0.2$& 10.8 &  
$0.22\pm0.04$&  $0.762\pm0.021$&  \nodata\nl
\nodata&  \nodata&  \nodata&  \nodata&    1.3 &  $ 1.8\pm0.2$& 10.2 &  
$0.06\pm0.04$&  $0.816\pm0.022$&  \nodata\nl
28&  980928&  51084.34&  C &    2.69 &  $13.9\pm0.2$&  11.2 &  
$-0.02\pm0.01$&  $0.970\pm0.003$&  \nodata\nl
\nodata&  \nodata&  \nodata&  \nodata&    5.4 &  $ 5.0\pm0.2$&  8.1 &  
$0.22\pm0.03$&  $0.709\pm0.020$&  \nodata\nl
\nodata&  \nodata&  \nodata&  \nodata&    1.3 &  $ 2.6\pm0.3$&  8.9 &  
$0.00\pm0.05$&  $0.741\pm0.026$&  \nodata\nl
29&  980929&  51085.27&  C &  4.13 &  $11.8\pm0.2$\tablenotemark{j} & 12.9 &  
$-0.08\pm0.01$&  $0.962\pm0.003$&  \nodata\nl
\nodata&  \nodata&  \nodata&  \nodata&    8.3 &  $ 3.2\pm0.1$& 10.0&  
$0.26\pm0.04$&  $0.693\pm0.025$&  \nodata\nl
\nodata&  \nodata&  \nodata&  \nodata&    2.1 &  $ 1.8\pm0.1$& 8.6 &  
$-0.10\pm0.04$&  $0.835\pm0.017$&  \nodata\nl
30&  980929&  51085.92&  C &    2.89 &  $14.0\pm0.3$&   9.9&  
$-0.04\pm0.02$&  $0.975\pm0.004$&  \nodata\nl
\nodata&  \nodata&  \nodata&  \nodata&    5.8 &  $ 3.9\pm0.2$&  13.4 &  
$0.22\pm0.05$&  $0.796\pm0.032$&  \nodata\nl
\nodata&  \nodata&  \nodata&  \nodata&    1.4 &  $ 2.6\pm0.3$&  6.5 &  
$0.03\pm0.06$&  $0.770\pm0.038$&  \nodata\nl
31&  980929&  51085.99&  C &  3.09 &  $13.3\pm0.2$\tablenotemark{j} &  7.5&  
$-0.04\pm0.01$&  $0.951\pm0.004$&  \nodata\nl
\nodata&  \nodata&  \nodata& \nodata& 6.1 & $ 4.1\pm0.1$\tablenotemark{j} & 8.0 &  
$0.18\pm0.03$&  $0.730\pm0.021$&  \nodata\nl
\nodata&  \nodata&  \nodata&  \nodata&  1.5 & $ 2.8\pm0.2$\tablenotemark{j} & 6.2 &  
$0.02\pm0.04$&  $0.775\pm0.023$&  \nodata\nl
32&  980930&  51086.89&  C &    3.51 &  $12.5\pm0.1$\tablenotemark{j} &  10.6 &  
$-0.06\pm0.01$&  $0.961\pm0.002$&  \nodata\nl
\nodata&  \nodata&  \nodata&  \nodata&    7.0 &  $ 3.7\pm0.1$& 10.5&  
$0.20\pm0.02$&  $0.727\pm0.016$&  \nodata\nl
\nodata&  \nodata&  \nodata&  \nodata&    1.8 &  $ 3.0\pm0.1$&  6.3 &  
$0.02\pm0.02$&  $0.798\pm0.014$&  \nodata\nl
33&  981001&  51087.72&  C &    3.44 &  $12.9\pm0.2$\tablenotemark{j} &  11.1 &  
$-0.05\pm0.01$&  $0.964\pm0.002$&  \nodata\nl
\nodata&  \nodata&  \nodata&  \nodata&    6.9 & $ 3.8\pm0.1$&  9.9 &  
$0.20\pm0.02$&  $0.765\pm0.017$&  \nodata\nl
\nodata&  \nodata&  \nodata&  \nodata&    1.7 & $ 3.4\pm0.2$&  6.8 &  
$0.02\pm0.03$&  $0.816\pm0.015$&  \nodata\nl
34&  981002&  51088.01&  C &    3.21 &  $13.3\pm0.2$&  10.7 &  
$-0.06\pm0.01$&  $0.968\pm0.003$&  \nodata\nl
\nodata&  \nodata&  \nodata&  \nodata&    6.4 &  $ 4.0\pm0.2$& 10.3 &  
$0.19\pm0.03$&  $0.775\pm0.020$&  \nodata\nl
\nodata&  \nodata&  \nodata&  \nodata&    1.6 &  $ 3.2\pm0.2$& 6.4 &  
$0.01\pm0.03$&  $0.812\pm0.017$&  \nodata\nl
35&  981003&  51089.01&  C &    3.04 &  $14.1\pm0.2$&   9.2&  
$-0.06\pm0.01$&  $0.975\pm0.003$&  \nodata\nl
\nodata&  \nodata&  \nodata&  \nodata&    6.1&  $ 3.3\pm0.2$&  15.2&  
$0.21\pm0.04$&  $0.784\pm0.025$&  \nodata\nl
\nodata&  \nodata&  \nodata&  \nodata&    1.5&  $ 3.3\pm0.3$&   6.9&  
$0.05\pm0.04$&  $0.801\pm0.026$&  \nodata\nl
36&  981004&  51090.14&  C &    3.93 &  $12.4\pm0.2$&  12.3 &  
$-0.08\pm0.01$&  $0.965\pm0.003$&  \nodata\nl
\nodata&  \nodata&  \nodata&  \nodata&    7.9 &  $ 3.5\pm0.2$& 10.9 &  
$0.20\pm0.05$&  $0.784\pm0.041$&  \nodata\nl
\nodata&  \nodata&  \nodata&  \nodata&    2.0 &  $ 3.4\pm0.2$&  5.8 &  
$-0.02\pm0.03$&  $0.849\pm0.019$&  \nodata\nl
37&  981004&  51090.70&  C &    3.72 &  $12.6\pm0.2$&  11.6 &  
$-0.08\pm0.01$&  $0.963\pm0.003$&  \nodata\nl
\nodata&  \nodata&  \nodata&  \nodata&    7.4 &  $ 3.8\pm0.1$& 9.8 &  
$0.19\pm0.04$&  $0.700\pm0.028$&  \nodata\nl
\nodata&  \nodata&  \nodata&  \nodata&    1.9 &  $ 3.6\pm0.2$& 5.5 &  
$-0.11\pm0.04$&  $0.843\pm0.020$&  \nodata\nl
38&  981005&  51091.74&  C &    5.60 &  $10.3\pm0.2$&  10.2 &  
$-0.15\pm0.02$&  $0.939\pm0.006$&  \nodata\nl
\nodata&  \nodata&  \nodata&  \nodata&   11.2 &  $ 2.2\pm0.1$& 12.9 &  
$0.26\pm0.07$&  $0.605\pm0.078$&  \nodata\nl
\nodata&  \nodata&  \nodata&  \nodata&    2.8 &  $ 3.7\pm0.2$&  3.8 &  
$-0.14\pm0.05$&  $0.881\pm0.029$&  \nodata\nl
39&  981007&  51093.14&  C &    6.55 &  $ 7.6\pm0.1$&  10.6 &  
$-0.23\pm0.02$&  $0.896\pm0.008$&  \nodata\nl
\nodata&  \nodata&  \nodata&  \nodata&   13.1 &  $ 1.3\pm0.2$& 11.5&  
$0.29\pm0.08$&  $0.477\pm0.050$&  \nodata\nl
\nodata&  \nodata&  \nodata&  \nodata&    3.3 &  $ 2.4\pm0.2$&  5.6 &  
$-0.23\pm0.04$&  $0.827\pm0.021$&  \nodata\nl
40&  981008&  51094.14&  C &    4.32 &  $12.0\pm0.2$&  12.7 &  
$-0.10\pm0.02$&  $0.965\pm0.004$&  \nodata\nl
\nodata&  \nodata&  \nodata&  \nodata&    8.6 &  $ 3.4\pm0.2$& 12.1 &  
$0.22\pm0.05$&  $0.780\pm0.061$&  \nodata\nl
\nodata&  \nodata&  \nodata&  \nodata&    2.2 &  $ 3.8\pm0.2$&  5.1 &  
$-0.04\pm0.05$&  $0.848\pm0.029$&  \nodata\nl
41&  981008&  51094.57&  C &    5.07 &  $10.7\pm0.2$&  12.7 &  
$-0.16\pm0.02$&  $0.950\pm0.004$&  \nodata\nl
\nodata&  \nodata&  \nodata&  \nodata&   10.1 &  $ 2.7\pm0.2$& 12.1 &  
$0.23\pm0.05$&  $0.613\pm0.038$&  \nodata\nl
\nodata&  \nodata&  \nodata&  \nodata&    2.5 &  $ 3.7\pm0.2$&  5.1 &  
$-0.05\pm0.04$&  $0.823\pm0.031$&  \nodata\nl
42&  981009&  51095.61&  C &    4.49 &  $11.9\pm0.3$&  12.5 &  
$-0.13\pm0.02$&  $0.96\pm0.01$&  \nodata\nl
\nodata&  \nodata&  \nodata&  \nodata&    8.9 &  $ 4.1\pm0.2$& 7.7 &  
$0.27\pm0.07$&  $0.73\pm0.06$&  \nodata\nl
\nodata&  \nodata&  \nodata&  \nodata&    2.2 &  $ 3.5\pm0.2$& 7.5 &  
$0.00\pm0.05$&  $0.83\pm0.03$&  \nodata\nl
43&  981010&  51096.57&  C &  5.40 &  $11.7\pm0.2$\tablenotemark{j} & 4.0 &  
$-0.12\pm0.02$&  $0.91\pm0.01$&  \nodata\nl
\nodata&  \nodata&  \nodata& \nodata& 10.2 &  $ 4.9\pm0.2$\tablenotemark{j} & 4.0 &  
$0.26\pm0.05$&  $0.60\pm0.03$&  \nodata\nl
\nodata&  \nodata&  \nodata& \nodata&  2.8 &  $ 4.9\pm0.2$\tablenotemark{j} & 4.0 &  
$-0.10\pm0.03$&  $0.84\pm0.02$&  \nodata\nl
44&  981011&  51097.57&  C &    4.74 &  $11.2\pm0.3$&  11.6 &  
$-0.13\pm0.02$&  $0.95\pm0.01$&  \nodata\nl
\nodata&  \nodata&  \nodata&  \nodata&    9.5 &  $ 3.0\pm0.2$& 11.0 &  
$0.39\pm0.07$&  $0.74\pm0.06$&  \nodata\nl
\nodata&  \nodata&  \nodata&  \nodata&    2.4 &  $ 4.1\pm0.2$&  5.5 &  
$-0.04\pm0.05$&  $0.92\pm0.02$&  \nodata\nl
45&  981011&  51097.81&  C &    4.20 &  $12.7\pm0.3$&  11.4 &  
$-0.14\pm0.03$&  $0.96\pm0.01$&  \nodata\nl
\nodata&  \nodata&  \nodata&  \nodata&    8.4 &  $ 3.8\pm0.3$& 11.1 &  
$0.33\pm0.08$&  $0.82\pm0.07$&  \nodata\nl
\nodata&  \nodata&  \nodata&  \nodata&    2.1 &  $ 3.8\pm0.3$&  7.0 &  
$0.05\pm0.08$&  $0.78\pm0.04$&  \nodata\nl
46&  981012&  51098.28&  C &    5.00 &  $12.0\pm0.2$&  7.5 &  
$-0.17\pm0.03$&  $0.96\pm0.01$&  \nodata\nl
\nodata&  \nodata&  \nodata&  \nodata&   10.0 &  $ 4.6\pm0.2$&  5.2 & 
$0.21\pm0.06$&  $0.77\pm0.06$&  \nodata\nl
\nodata&  \nodata&  \nodata&  \nodata&    2.5 &  $ 5.5\pm0.2$&  4.4 &  
$-0.10\pm0.06$&  $0.85\pm0.03$&  \nodata\nl
47&  981013&  51099.21&  C &    4.85 &  $11.0\pm0.2$& 11.8 &  
$-0.15\pm0.02$&  $0.95\pm0.11$&  \nodata\nl
\nodata&  \nodata&  \nodata&  \nodata&    9.7 &  $ 3.0\pm0.2$& 11.5 &  
$0.52\pm0.08$&  $0.74\pm0.18$&  \nodata\nl
\nodata&  \nodata&  \nodata&  \nodata&    2.4 &  $ 4.3\pm0.2$&  5.4 &  
$-0.15\pm0.06$&  $0.82\pm0.15$&  \nodata\nl
48&  981013&  51099.61&  C &    4.99 &  $10.8\pm0.2$& 10.6 &  
$-0.18\pm0.02$&  $0.95\pm0.01$&  \nodata\nl
\nodata&  \nodata&  \nodata&  \nodata&   10.0 &  $ 3.6\pm0.1$&  7.4 &  
$0.27\pm0.06$&  $0.85\pm0.08$&  \nodata\nl
\nodata&  \nodata&  \nodata&  \nodata&    2.5 &  $ 3.3\pm0.2$&  7.6 &  
$-0.02\pm0.04$&  $0.82\pm0.02$&  \nodata\nl
49&  981014&  51100.29&  C &    6.48 &  $ 7.3\pm0.2$& 12.0 &  
$-0.29\pm0.03$&  $0.90\pm0.01$&  \nodata\nl
\nodata&  \nodata&  \nodata&  \nodata&   13.0 &  $ 1.5\pm0.1$& 13.0 &  
$0.34\pm0.09$&  $0.59\pm0.08$&  \nodata\nl
\nodata&  \nodata&  \nodata&  \nodata&    3.2 &  $ 3.5\pm0.2$&  4.9 &  
$-0.15\pm0.05$&  $0.83\pm0.04$&  \nodata\nl
50&  981015&  51101.61&  C &    6.85 &  $ 6.4\pm0.1$&  7.7 &  
$-0.28\pm0.04$&  $0.83\pm0.02$&  $141\pm3$\nl
\nodata&  \nodata&  \nodata&  \nodata&   13.7 &  $ 1.2\pm0.1$& 11.1 &  
$0.10\pm0.10$&  $0.61\pm0.11$&  \nodata\nl
\nodata&  \nodata&  \nodata&  \nodata&    3.4 &  $ 3.1\pm0.1$&  4.9 &  
$-0.25\pm0.07$&  $0.82\pm0.04$&  \nodata\nl
51&  981015&  51101.94&  C &    6.77 &  $ 7.5\pm0.1$&  7.0 &  
$-0.34\pm0.04$&  $0.89\pm0.01$&  $145\pm8$\nl
\nodata&  \nodata&  \nodata&  \nodata&   13.5 &  $ 1.7\pm0.1$& 7.7 &  
$0.30\pm0.09$&  $0.46\pm0.05$&  \nodata\nl
\nodata&  \nodata&  \nodata&  \nodata&    3.4 &  $ 4.5\pm0.1$& 3.3 &  
$-0.31\pm0.06$&  $0.78\pm0.03$&  \nodata\nl
52&  981020&  51106.95&  B&    5.46 &  $ 3.5\pm0.1$& 10.7 & 
$0.24\pm0.04$&  $0.99\pm0.03$&  $194\pm15$\nl
\nodata&  \nodata&  \nodata&  \nodata&   10.9 &  $ 1.5\pm0.1$&  7.7 &  
$-0.54\pm0.14$&  $1.28\pm0.34$&  \nodata\nl
\nodata&  \nodata&  \nodata&  \nodata&    2.7 &  $ 0.7\pm0.1$&  5.1 &  
$-0.53\pm0.36$&  $0.0\pm1$&  \nodata\nl
53&  981022&  51108.08&  B&    5.4 &  $ 4.0\pm0.2$\tablenotemark{k} & 3.2 & 
$0.27\pm0.02$&  $0.98\pm0.01$&  $183\pm3$\nl
\nodata&  \nodata&  \nodata&  \nodata&   10.9&  $ 2.2\pm0.1$&   6.0 &  
$-0.39\pm0.04$&  $0.97\pm0.05$&  \nodata\nl
\nodata&  \nodata&  \nodata&  \nodata&    2.7&  $ 1.5\pm0.1$&   3.8 &  
$-0.35\pm0.12$&  $0.70\pm0.22$&  \nodata\nl
54&  981023&  51109.74&  B&    4.94 &  $ 3.8\pm0.1$& 12.7 & 
$0.13\pm0.03$&  $0.99\pm0.02$&  \nodata\nl
\nodata&  \nodata&  \nodata&  \nodata&    9.9 &  $ 1.6\pm0.1$& 8.3 &  
$-0.27\pm0.15$&  $0.99\pm0.26$&  \nodata\nl
59&  981029&  51115.28&  A&    6.89 &  $ 2.0\pm0.1$&   2.8 &  
$-1.45\pm0.51$&  $0.1\pm1$&  $270\pm6$ \tablenotemark{l}\nl
\nodata&  \nodata&  \nodata&  \nodata&   16.6 &  $0.53\pm0.06$&  8.3 &  
$0.22\pm0.59$&  $0.0\pm1$&  \nodata\nl
64&  981109&  51126.59& A?  &    4.88 &  $ 1.5\pm0.1$&   4.9 &  
$-1.78\pm1.20$&  $0.2\pm1$&  $284\pm4$ \tablenotemark{l} \nl
151&  990302&  51239.08&  ?&   18.1 &  $ 0.50\pm0.03$&  18.9 &
  $-0.56\pm0.19$ &  $0.9\pm1$ &  \nodata\nl
153&  990304&  51241.83&  A&    5.84 &  $ 2.0\pm0.1$&  5.8 &  
$-0.75\pm0.11$&  $1.3\pm1$& $283\pm2$ \tablenotemark{l}\nl
154&  990305&  51242.51&  A&    5.62 &  $ 1.3\pm0.1$&  5.6 &  
$-1.34\pm0.32$&  $0.4\pm1$& $283\pm3$ \tablenotemark{l}\nl
155&  990307&  51244.50& A?  &    8.45 &  $ 1.3\pm0.1$&  2.2 &  
$-0.76\pm0.39$&  $0.0\pm1$&  \nodata\nl
156&  990308&  51245.35& B& 6.38 & $ 3.3\pm0.1$\tablenotemark{j} & 13.9 &  
$0.34\pm0.03$&  $0.96\pm0.01$&  $182\pm2$\nl
\nodata&  \nodata& \nodata& \nodata& 12.2 & $ 1.3\pm0.1$\tablenotemark{j} & 5.6 &  
$-0.44\pm0.08$&  $1.03\pm0.13$&  \nodata\nl
\nodata&  \nodata&  \nodata& \nodata& 3.1 & $ 1.5\pm0.1$\tablenotemark{j} & 4.0 &  
$-0.63\pm0.24$&  $1.5\pm1$&  \nodata\nl
157&  990309&  51246.41& A?  &  7.70 &  $ 0.9\pm0.1$\tablenotemark{j} & 4.3 &  
$-0.14\pm0.35$&  $1.0\pm1$&  $214\pm7$\nl
158&  990310&  51247.98&  B&  6.15 &  $ 3.4\pm0.1$\tablenotemark{j} & 11.8 &  
$0.33\pm0.03$&  $0.97\pm0.01$& $189\pm6$\nl
\nodata&  \nodata&  \nodata&  \nodata&    3.1 &  $ 1.6\pm0.1$&  6.2 &  
$-0.49\pm0.08$&  $0.78\pm0.08$&  \nodata\nl
159&  990311&  51248.09&  B&    5.89 &  $ 3.5\pm0.1$& 12.3 &  
$-0.50\pm0.15$&  $0.63\pm0.11$&  \nodata\nl
\nodata&  \nodata&  \nodata&  \nodata&   12.3 &  $ 1.3\pm0.1$&  9.3 &  
$0.18\pm0.04$&  $1.00\pm0.02$&  \nodata\nl
\nodata&  \nodata&  \nodata&  \nodata&   11.8 &  $ 1.3\pm0.1$& 8.5 &  
$-0.38\pm0.14$&  $1.06\pm0.30$&  \nodata\nl
\nodata&  \nodata&  \nodata&  \nodata&    2.9 &  $ 0.8\pm0.1$& 6.1 &  
$-0.74\pm0.43$&  $0.6\pm1$&  \nodata\nl
160&  990312&  51249.40&  B&  6.1 &  $ 3.8\pm0.1$\tablenotemark{j} & 9.4 &  
$0.34\pm0.06$&  $0.99\pm0.03$&  $185\pm18$\nl
\nodata&  \nodata&  \nodata&  \nodata&   12.2 &  $ 1.9\pm0.1$&  5.7 &  
$-0.47\pm0.14$&  $1.41\pm0.41$&  \nodata\nl
\nodata&  \nodata&  \nodata&  \nodata&    3.1 &  $ 2.8\pm0.1$&  5.3 &  
$-0.23\pm0.17$&  $0.74\pm0.17$&  \nodata\nl
161&  990313&  51250.69&  C &    6.71 &  $ 6.2\pm0.1$&  9.4 &  
$-0.35\pm0.04$&  $0.89\pm0.03$&  $102\pm3$ \tablenotemark{l}\nl
\nodata&  \nodata&  \nodata&  \nodata&   13.4 &  $ 1.5\pm0.1$& 10.9 &  
$0.52\pm0.15$&  $0.51\pm0.15$&  \nodata\nl
\nodata&  \nodata&  \nodata&  \nodata&    3.4 &  $ 3.3\pm0.2$&  5.0 &  
$-0.08\pm0.07$&  $0.72\pm0.06$&  \nodata\nl
162&  990316&  51253.22&  B& 5.4 & $ 4.4\pm0.1$\tablenotemark{j} & 5.0 &  
$0.27\pm0.05$&  $1.00\pm0.03$&  \nodata\nl
\nodata&  \nodata&  \nodata& \nodata& 10.7 & $ 2.2\pm0.1$\tablenotemark{j} & 4.0 &  
$-0.47\pm0.11$&  $1.37\pm0.34$&  \nodata\nl
\nodata&  \nodata&  \nodata& \nodata& 2.9 &  $ 2.1\pm0.1$\tablenotemark{j} & 4.0 &  
$-0.18\pm0.38$&  $0.8\pm1$&  \nodata\nl
163&  990317&  51254.09&  B&    5.95 &  $ 3.0\pm0.1$&   5.9 &  
$0.08\pm0.08$&  $0.62\pm0.05$&  $209\pm6$ \nl
\nodata&  \nodata&  \nodata&  \nodata&    3.0 &  $ 1.8\pm0.1$& 6.2 &  
$-0.47\pm0.19$&  $0.28\pm0.05$&  \nodata\nl
164&  990318&  51255.09&  A&   10.1\tablenotemark{l} &  $ 8.0\pm0.2$&   1.2 &  
$-0.90\pm0.36$&  $0.4\pm1$&  $281\pm4$ \tablenotemark{l}\nl
\nodata&  \nodata&  \nodata&  \nodata&   13.6&  $ 1.8\pm0.1$&   2.1&  
$0.02\pm0.33$&  $1.0\pm1$&  \nodata\nl
166 &  990321&  51258.30& A?  &  10.3\tablenotemark{l} &  $ 6.9\pm0.2$\tablenotemark{l,k} &
   1.3 &  $-0.43\pm0.27$&  $1.0\pm1$&  $275\pm4$ \tablenotemark{l}\nl
178&  990402&  51270.74& A?  &    7.1\tablenotemark{l} &  $ 12.5\pm0.2$\tablenotemark{l,k} &
  $\sim1$ & $0.12\pm0.19$&  $1.7\pm1$&  $253\pm9$\nl
179&  990403&  51271.41& A?  &    7.1\tablenotemark{l} &  $ 12.5\pm0.2$\tablenotemark{l,k} &
  $\sim1$ & $-0.07\pm0.22$&  $1.0\pm1$&  \nodata\nl
\enddata

\tablenotetext{a}{Start of observation, $MJD=JD-2,400,000.5$.}
\tablenotetext{b}{QPO centroid frequency ; uncertainties are 
generally $<5$\% at the 95\% confidence level.}
\tablenotetext{c}{The QPO amplitude is the fractional rms 
fluctuation, calculated as the square root of the integrated power 
in the QPO feature and expressed as a fraction of the mean
count rate.  Errors are $1\sigma$, with a lower limit of 0.1\% 
for systematic uncertainty in the continuum model, 
except for especially narrow QPOs ($Q > 10$).}
\tablenotetext{d}{Q = QPO Frequency/FWHM}
\tablenotetext{e}{Phase lag between 2--13~keV and 13--30~keV bands 
at the QPO center.  A positive value corresponds to a hard lag.  
Errors are $1\sigma$.}
\tablenotetext{f}{Coherence between the two energy bands at the QPO
center. Errors are $1\sigma$.}
\tablenotetext{g}{High-frequency QPOs detected with $4 \sigma$ 
level of confidence, nominally over the full energy range of the 
PCA instrument.}
\tablenotetext{h}{Special cases (noted in col. 1). soft band: 
2--11.3~keV; hard band 11.3--18.3~keV.}
\tablenotetext{i}{Special cases (noted in col. 1). soft band: 
2--6.5~keV; hard band: 6.5--30~keV.}
\tablenotetext{j}{QPO wings deviate from Lorentzian profile; QPO width
and amplitude may be slightly underestimated.} 
\tablenotetext{k}{QPO profile is not a Lorentzian; amplitude and 
width are calculated numerically.} 
\tablenotetext{l}{QPOs that are more clearly evident and fit by
using only PCA detections above 6 keV.}

\end{deluxetable}

\begin{deluxetable}{lccc}
\tablenum{2}
\tablecaption{Summary of QPO Types}
\tablehead{
 \colhead{Property} & \colhead{Type A\tablenotemark{a}} & 
 \colhead{Type B} & \colhead{Type C}
}
\startdata
Frequency (Hz) &  $\sim6$ &  5--6 &  0.1--10 \\
Amplitude (\%rms)&  3--4&  $\sim4$&  3--16 \\
$Q~~(\nu / FWHM) $ &  $\sim$2--4&  $\sim4$&  $\ga10$ \\
Phase Lag (rad.)&  $-0.6$ to $-1.4$&  0 to 0.4&  0.05 to $-0.4$ \\
~~Sub-Harmonic& \nodata &  soft &  soft \\
~~1st Harmonic&  soft &  soft &  hard \\
Coherence &  $<0.5$&  $\sim1$&  $\sim0.9$ \\
HFQPO\tablenotemark{b} &  4/4 &  6/9 &  5/51 \\
\enddata

\tablenotetext{a} {This column excludes the 6 ``A?'' type QPOs, which have broad profiles but statistically uncertain measurements of the coherence function.}
\tablenotetext{b} {In addition to these, HFQPOs are seen in one peculiar type and in 4 of 6 ``A?'' type of LFQPO.}

\end{deluxetable}

\figcaption[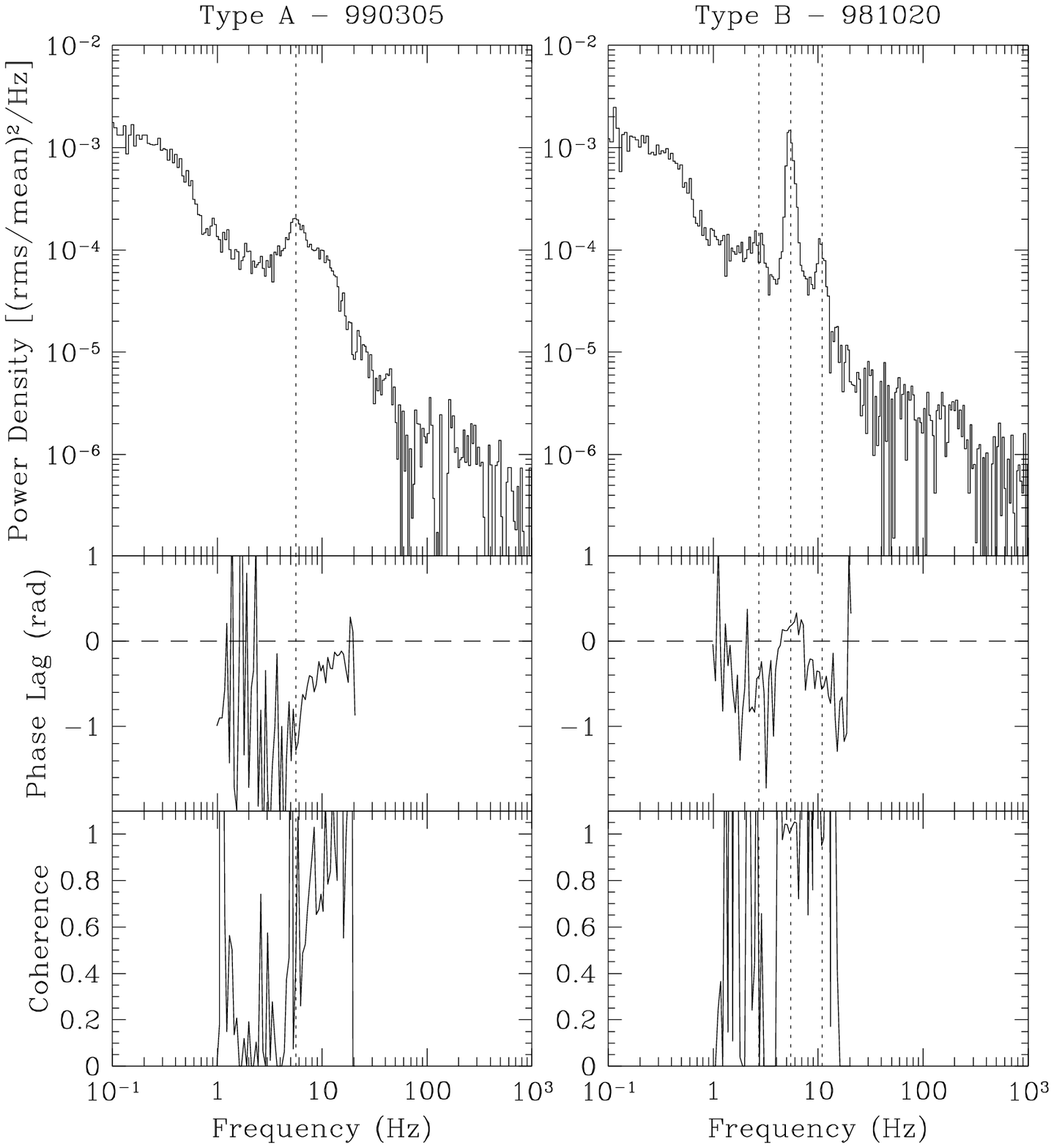]{Characteristic power spectrum (top), phase lag
spectrum (middle), and coherence function (bottom) for Type~A QPOs
(left panels) and Type~B QPOs (right panels).  The phase lag and coherence are
computed between the 2--13~keV and 13--30~keV bands, with a positive
phase lag representing a hard lag.  \label{f1}}

\figcaption[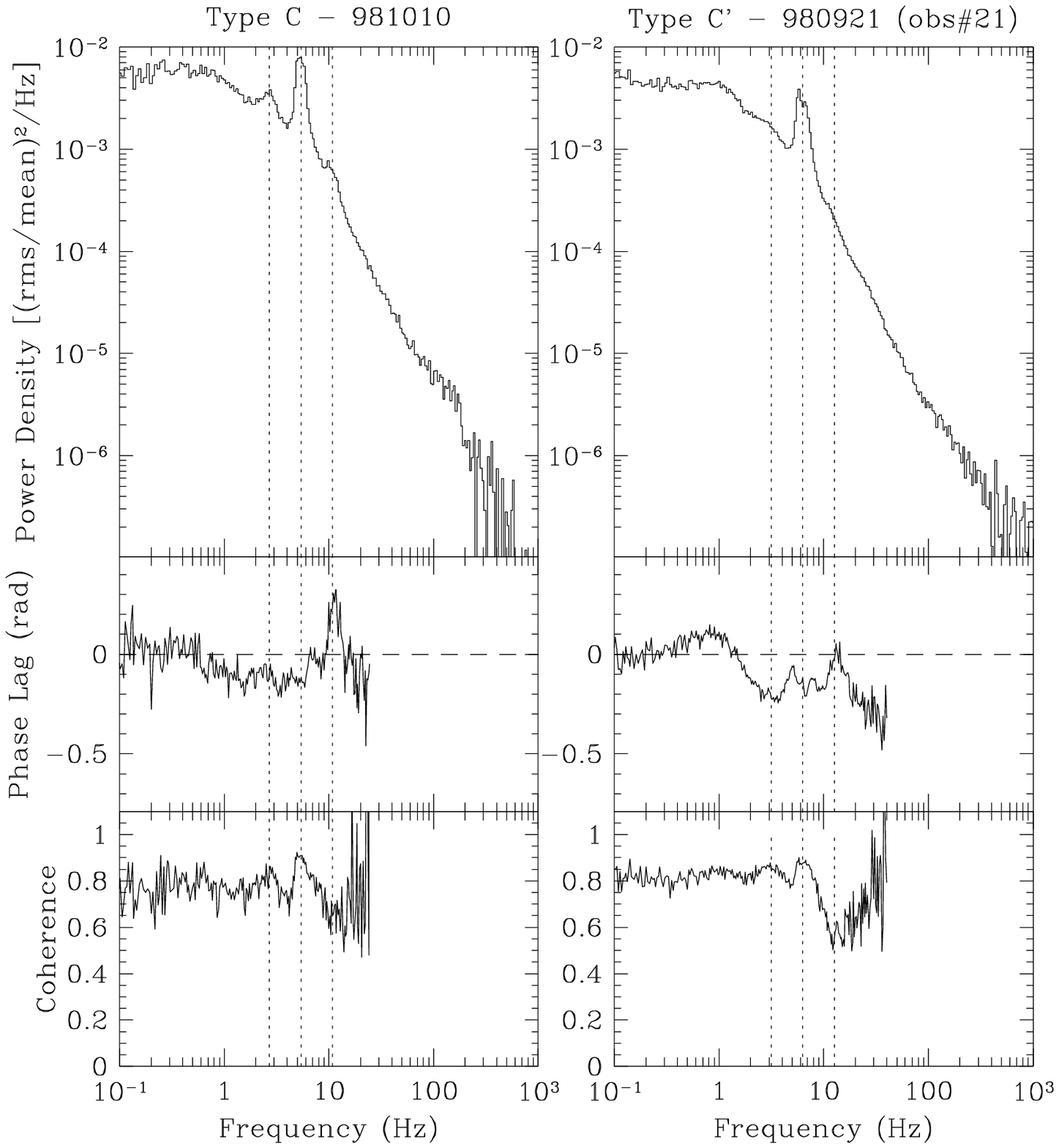]{Characteristic power spectrum (top), phase lag
spectrum (middle), and coherence function (bottom) for Type~C QPOs
(left panels) and Type C\'{ } QPOs (right panels). \label{f2}}

\figcaption[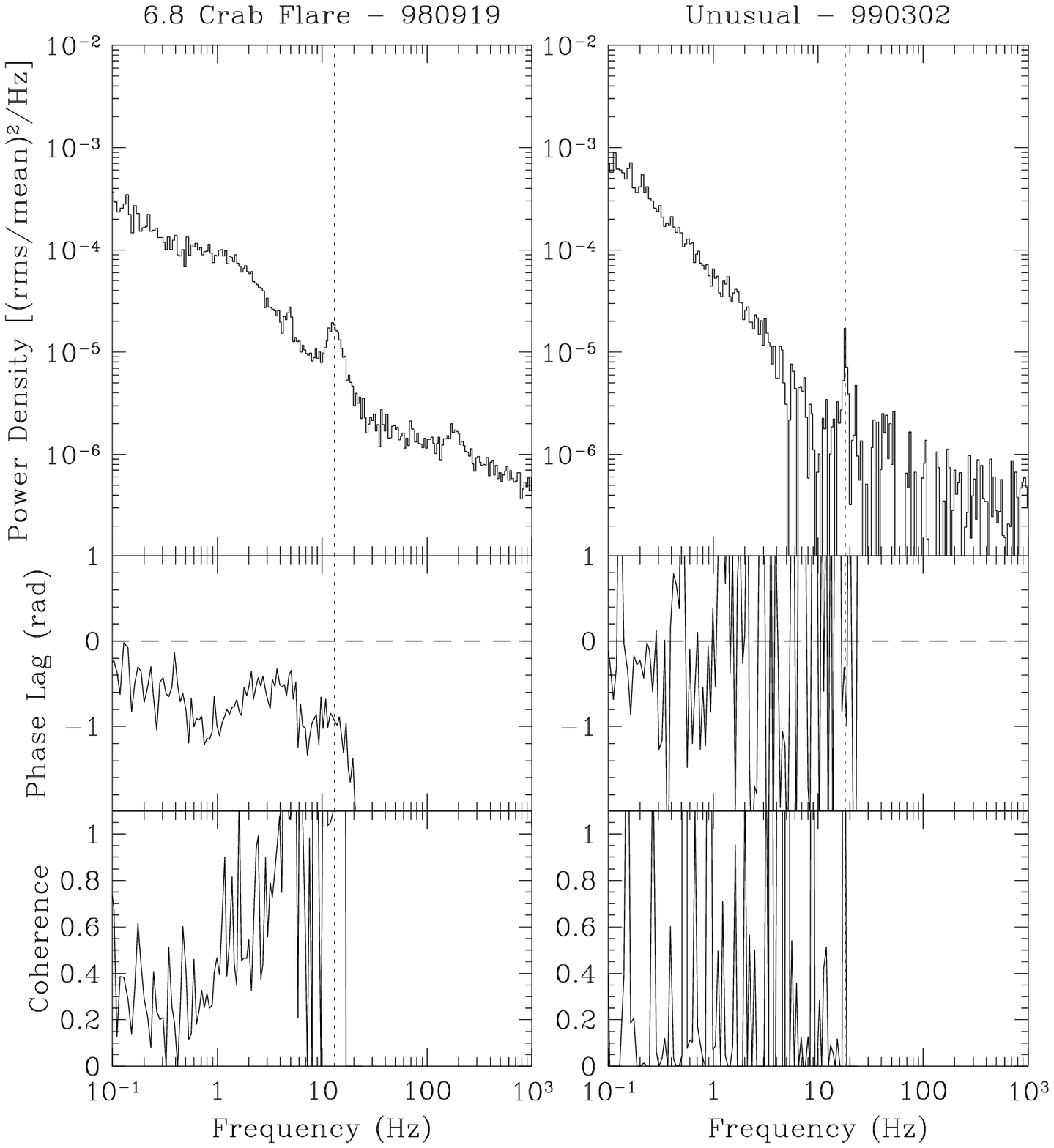]{Characteristic power spectrum (top), phase lag
spectrum (middle), and coherence function (bottom) for two anomalous
QPOs which do not resemble types A, B, or C (see text). \label{f3}}

\figcaption[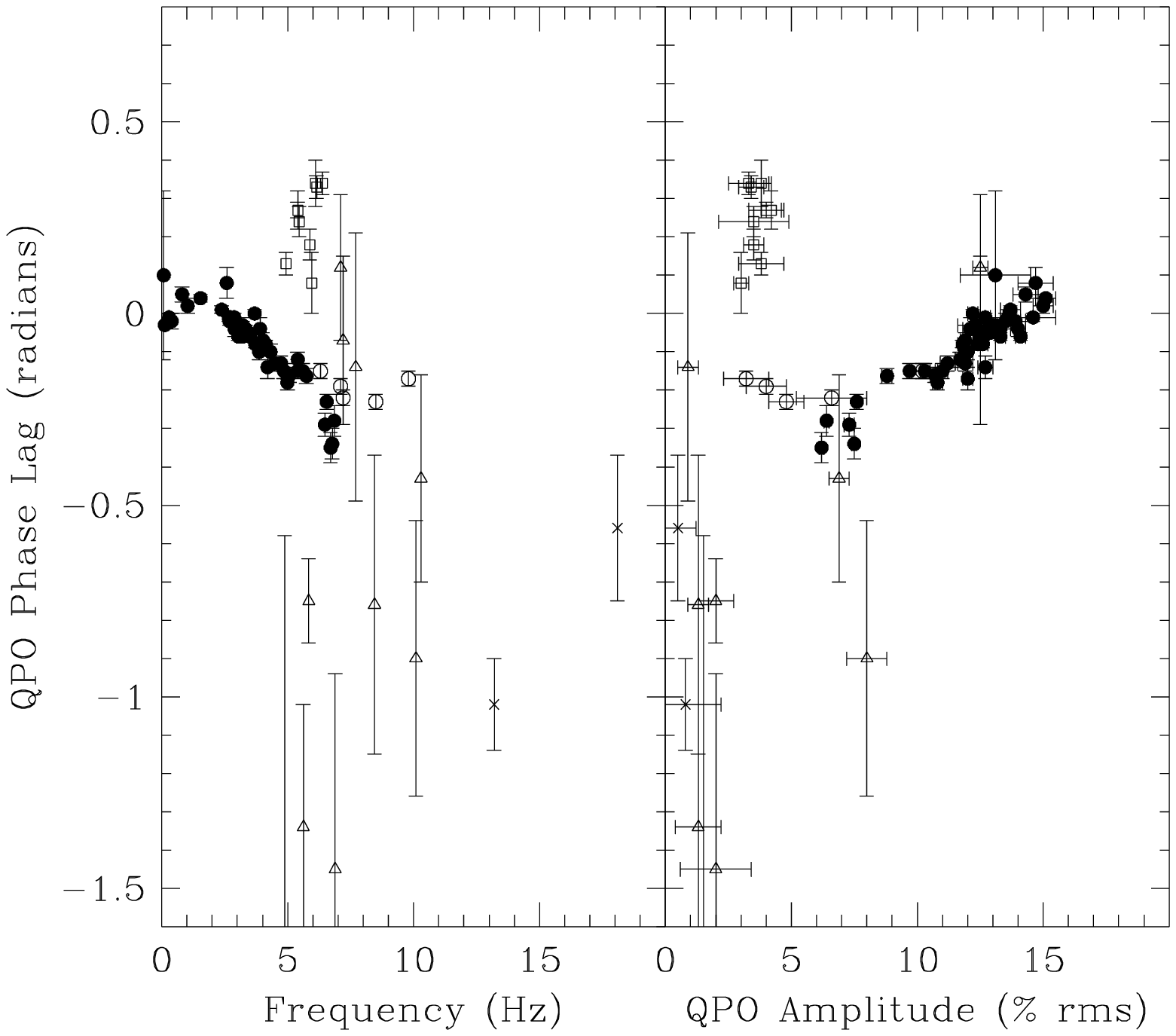]{Phase lag in the fundamental feature
vs. frequency (left panel) and integrated rms amplitude (right
panel). The plotting symbols distinguish the QPO type: Types A and A?
-- open triangles, Type B -- open squares, Type C -- filled circles,
Type C\'{ } -- open circles, and the anomalous -- `x'. \label{f4}}

\figcaption[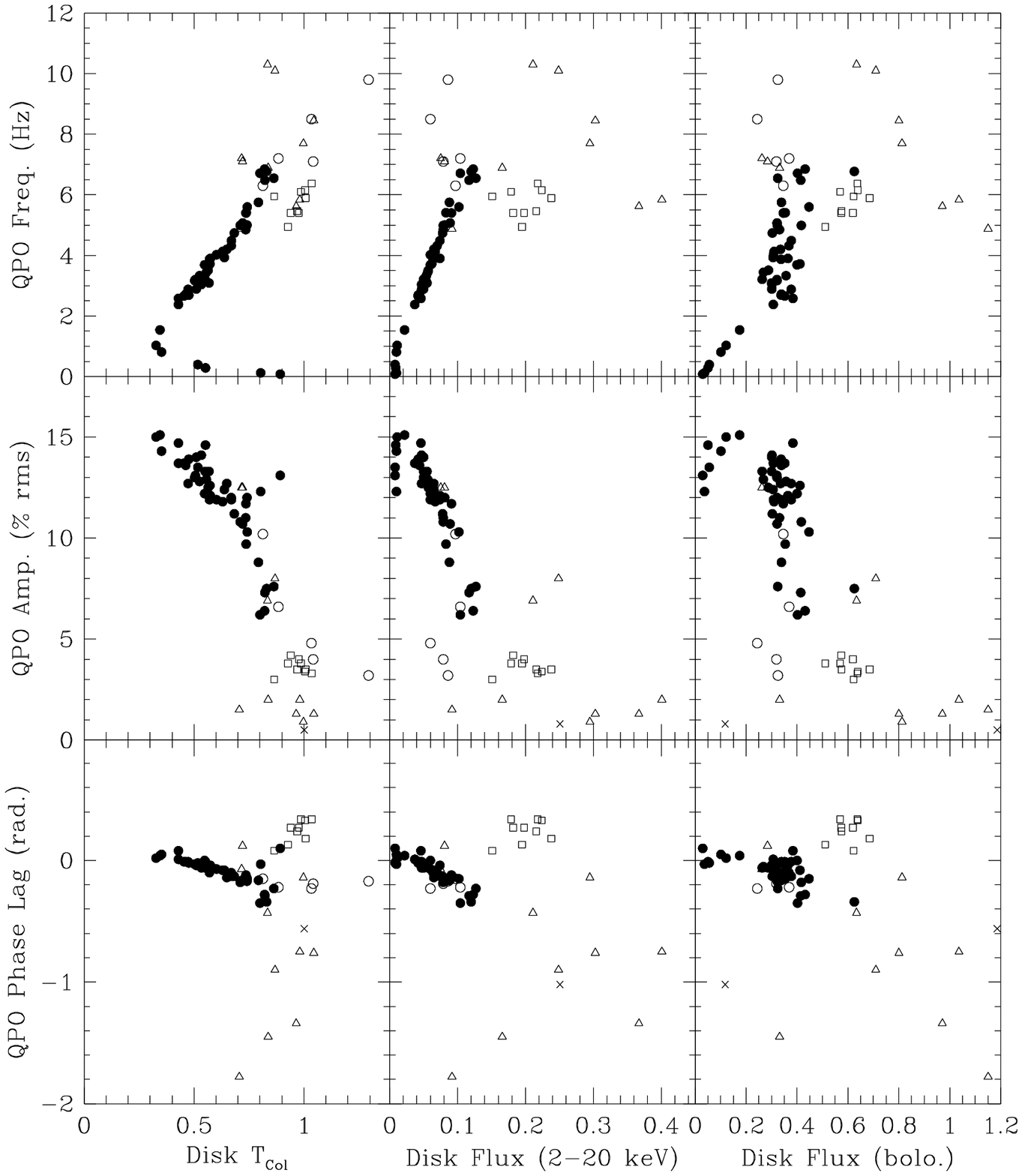]{QPO properties (fundamental feature) versus the
color temperature (keV) of the accretion disk (left panels), the
unabsorbed disk flux at 2--20 keV (center panels) and the bolometric
disk flux (right panels).  The fluxes are in units of $10^{-7}$ ergs
cm$^{-2}$ s$^{-1}$, and the plotting symbols are the same as in
Fig.~4. \label{f5}}

\figcaption[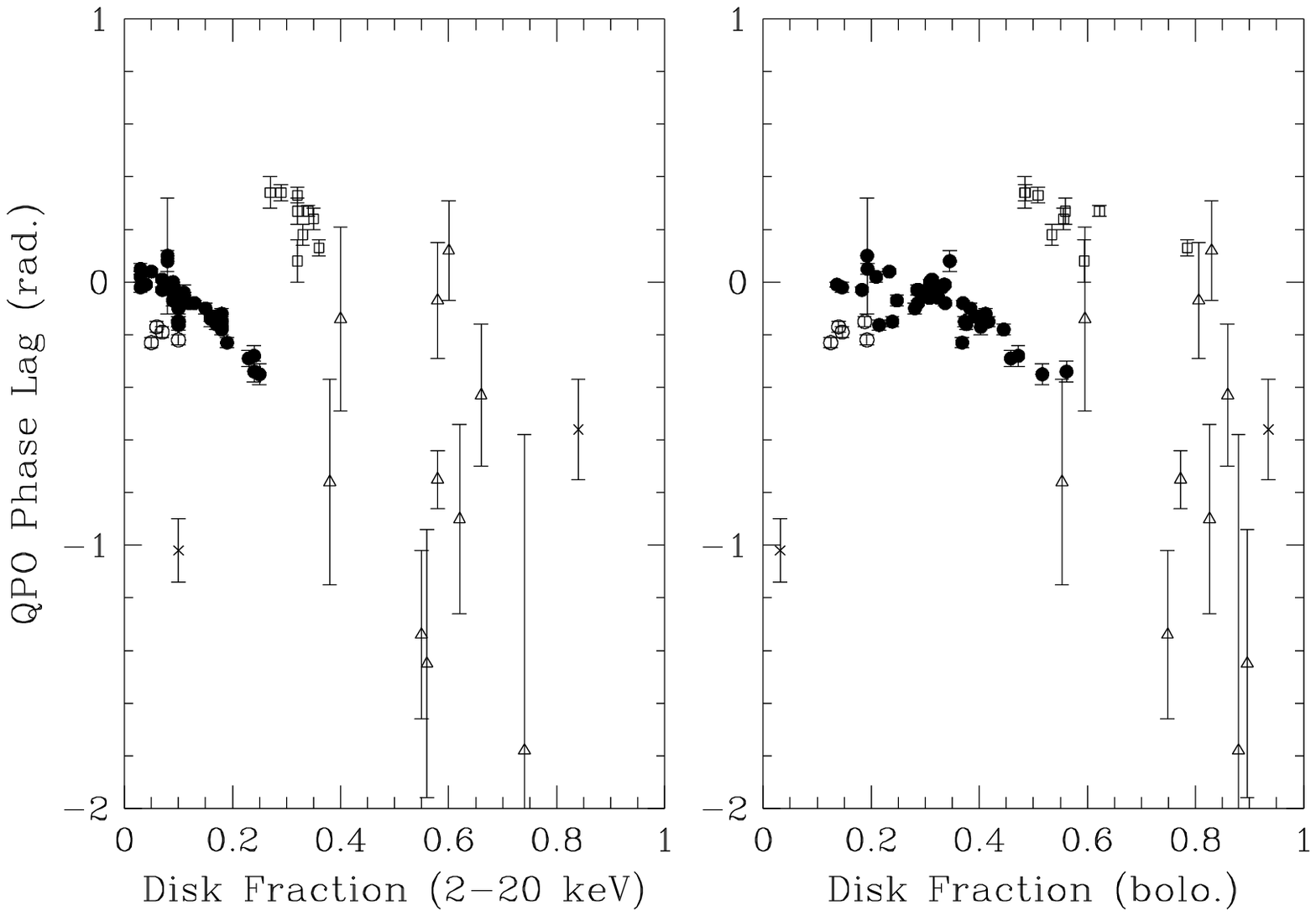]{QPO phase lag (fundamental feature) versus the
fraction of the unabsorbed, total flux attributed to the accretion
disk.  In the left panel, the integration of each spectral component
is limited to the range 2--20 keV. In the right panel we consider the
entire disk-blackbody spectrum, and we integrate the spectrum of the
power-law component over the range 1--30 keV. The latter is subject
to increased systematic uncertainty due to the extrapolation of the
spectrum well beyond the sensitivity band of the PCA instrument. The
plotting symbols are the same as in Fig.~4.
\label{f6}}

\figcaption[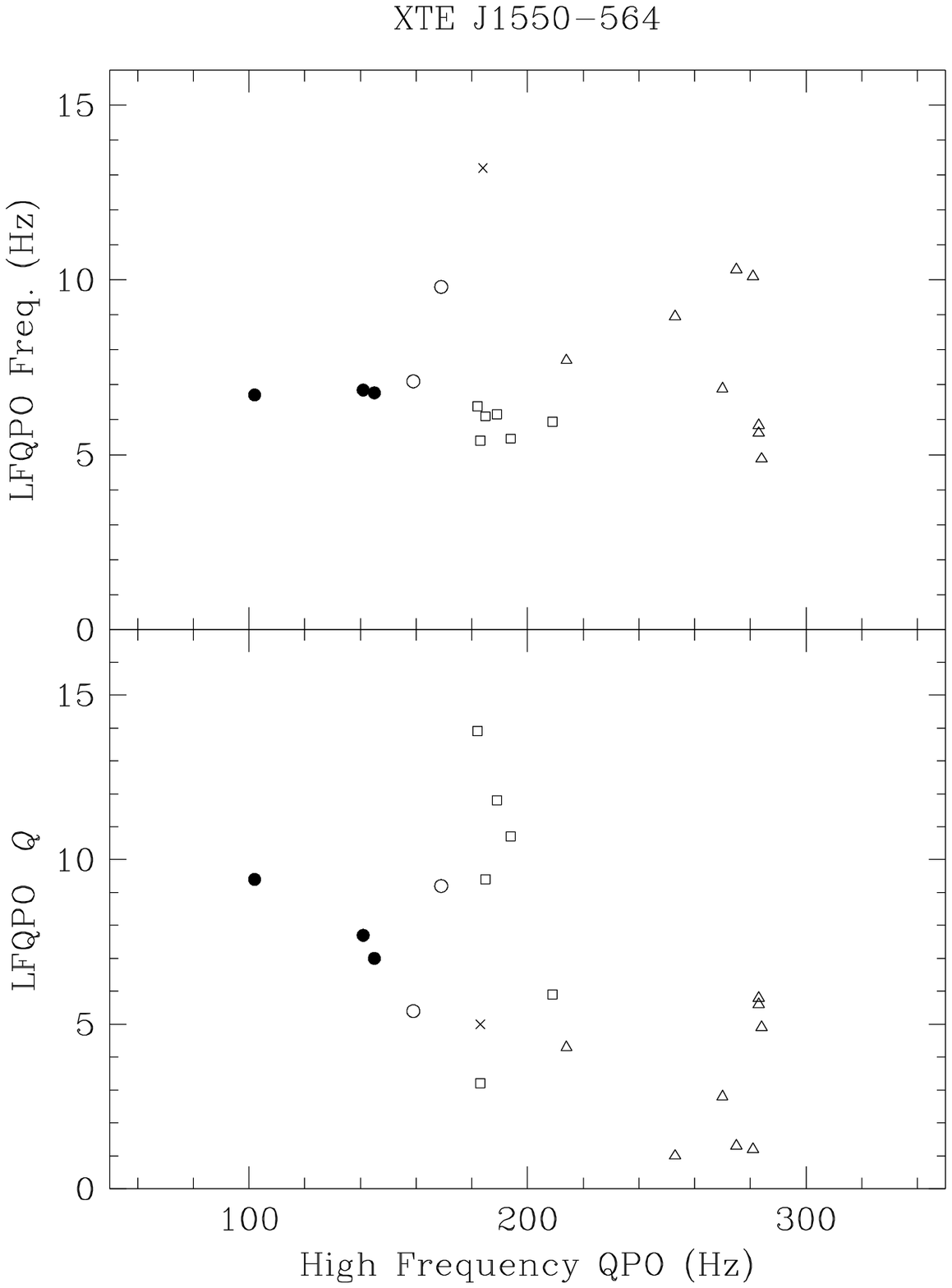]{The frequencies of HFQPOs in XTE|J1550--564
versus the LFQPO fundamental (top panel) and the fundamental's $Q$
value (bottom panel. The plotting symbols are the same as in Fig.~4.
The LFQPO type (but not the actual frequency) is correlated with the
frequency of the HFQPOs, while the fundamental's Q value decreases as
the HFQPOs progress toward narrow profiles and hard spectra when seen
at 284 Hz. \label{f7}}

\newpage
\begin{figure}
\figurenum{1}
\plotone{f1.eps}
\caption{ }
\end{figure}

\newpage
\begin{figure}
\figurenum{2}
\plotone{f2.eps}
\caption{ }
\end{figure}

\newpage
\begin{figure}
\figurenum{3}
\plotone{f3.eps}
\caption{ }
\end{figure}

\newpage
\begin{figure}
\figurenum{4}
\plotone{f4.eps}
\caption{ }
\end{figure}

\newpage
\begin{figure}
\figurenum{5}
\plotone{f5.eps}
\caption{ }
\end{figure}

\newpage
\begin{figure}
\figurenum{6}
\plotone{f6.eps}
\caption{ }
\end{figure}

\newpage
\begin{figure}
\figurenum{7}
\plotone{f7.eps}
\caption{ }
\end{figure}

\end{document}